\documentclass[12pt,showpacs,showkeys,amsmath,amssymb]{revtex4-1}
\usepackage{amsmath,amsfonts,amsthm,amscd,amssymb,latexsym}
\usepackage{bm}
\usepackage{dcolumn}
\usepackage{graphicx}
\usepackage{epstopdf}
\usepackage{color}
\usepackage{epsf}
\usepackage{epsfig}

\def\@{\partial_}

\def\negenspace{\kern-1.1em}

\def\sqr#1#2{{\vcenter{\hrule height.#2pt\hbox{\vrule width.#2pt
height#1pt \kern#1pt \vrule width.#2pt}\hrule height.#2pt}}}

\date{\today}

\begin{document}
\title{Spin Precession in Inertial and Gravitational Fields}

\author{Bahram Mashhoon}
\email{mashhoonb@missouri.edu}
\affiliation{Department of Physics and Astronomy,
University of Missouri, Columbia, Missouri 65211, USA}
\author{Yuri N. Obukhov}
\email{yo@thp.uni-koeln.de}
\affiliation{Theoretical Physics Laboratory, Nuclear Safety Institute, 
Russian Academy of Sciences, B.Tulskaya 52, 115191 Moscow, Russia}

\begin{abstract} 
We discuss the motion of spin in inertial and gravitational fields. The coupling of spin with rotation and the gravitomagnetic field has already been extensively studied; therefore, we focus here on the inertial and gravitational spin-orbit couplings. In particular, we investigate the classical and quantum aspects of spin precession and spin-orbit coupling in an arbitrary translationally accelerated frame of reference as well as the exterior Schwarzschild spacetime. Moreover, in connection with Einstein's principle of equivalence, we clarify the relation between the inertial and gravitational spin-orbit couplings. 
\end{abstract}



\maketitle

\section{Introduction}\label{intro}


Imagine a \emph{free} spinning test particle in spacetime. To begin with, we are interested in the motion of the particle spin as described by noninertial observers in Minkowski spacetime. Such observers are characterized by an antisymmetric acceleration tensor whose ``electric" and ``magnetic" components correspond respectively to the observer's 4-acceleration and the rotation of its spatial frame relative to local comoving nonrotating axes. It turns out that the spin couples \emph{differently} to the observer's translational acceleration (i.e., its 4-acceleration) than to the angular velocity of rotation of its actual spatial frame relative to a comoving nonrotating frame.  Analogous couplings are found in a gravitational field with respect to the natural tetrad frame of the fundamental observers at rest. Indeed, in the linear weak-field approximation, the couplings of particle spin to the gravitoelectric and gravitomagnetic fields are very similar to the corresponding couplings to the ``electric" and ``magnetic" components of the acceleration tensor of a noninertial observer in Minkowski spacetime.  It turns out that the couplings of the spin to rotation and the gravitomagnetic field are related by the gravitational Larmor theorem, which is a consequence of Einstein's principle of equivalence. This intimate relationship thus leads to the \emph{spin-rotation-gravity coupling} that has been extensively studied in both classical and quantum domains. On the other hand, the couplings of the spin to 4-acceleration and the gravitoelectric field, though very similar in form, are not \emph{directly} connected by a simple application of Einstein's heuristic principle of equivalence. Moreover, the connection between the classical and quantum results has not been clear in this case. The purpose of this paper is to clarify this confusing situation by filling in the gap in the calculations and then providing a consistent correspondence between the classical and quantum results. 

We consider spin precession within the framework of general theory of relativity. The same results are naturally expected in the teleparallel equivalent of general relativity and we explain the origin of any possible discrepancies between the two theories. 
In our convention,  the Minkowski metric tensor  $\eta_{\alpha \beta}$ is given by diag$(-1,1,1,1)$; moreover, Greek indices run from 0 to 3, while Latin indices run from 1 to 3. The hatted Greek indices ${\hat{\alpha}}$, ${\hat{\beta}}$, etc., refer to \emph{anholonomic} tetrad indices, while $\mu$, $\nu$, etc., refer to \emph{holonomic} spacetime indices. 


The motion of a free gyroscope in a gravitational field is described by the Mathisson-Papapetrou-Dixon equations~\cite{Ma, Pa, Di}; for recent studies of these equations, see~\cite{MS, PO}. Restricting our attention to a small but extended pole-dipole test particle and \emph{neglecting second-order spin effects}, it is possible to show that the spin vector $S^\mu$ in this case satisfies~\cite{CMP} 
\begin{equation}\label{I1}
 S_\mu u^\mu \approx 0\,, \qquad {\frac{DS^\mu}{d\tau}} \approx 0\,,
\end{equation}
where $u^\mu = dx^\mu/d\tau$ is the 4-velocity of a representative point (``center of mass") inside the gyroscope and $\tau$ is the proper time along this worldline. Moreover, 
\begin{equation}\label{I2}
 \frac{Du^\mu}{d\tau} \approx -\,{\frac{1}{2m}} R^{\mu}{}_{\nu \alpha \beta} u^\nu S^{\alpha \beta}\,,
\end{equation}
where $m$ is the mass of the gyroscope and the spin tensor is given by 
\begin{equation}\label{I3}
 S^{\mu \nu} \approx  {\frac 1c}\epsilon^{\mu \nu \rho \sigma} u_\rho S_\sigma\,.
\end{equation}
Here,  $\epsilon_{\mu \nu \rho \sigma}$ is the Levi-Civita tensor. The Mathisson-Papapetrou equations---when supplemented with the Frenkel-Pirani condition---have another principal interpretation that involves the motion of a free classical point particle with  ``intrinsic" spin~\cite{Pi, Ta, M0}. In this case, it can be shown that the ``intrinsic" spin vector $S^\mu$ is Fermi-Walker transported along the path of the point ``gyroscope" with 4-velocity $u^\mu$~\cite{Pi}. In this interpretation, it is noteworthy that Eqs.~\eqref{I1}-\eqref{I3} turn out to be valid as well when second-order spin effects can be neglected. 

It is clear from Eq.~\eqref{I2} that spin couples to spacetime curvature thereby giving rise to the Mathisson force~\cite{Ma}; therefore, Einstein's local principle of equivalence may not be generally applicable to spin precession.  Henceforth, we will work to linear order in spin, and so, for the sake of simplicity, we will drop the approximate equality signs in Eqs.~\eqref{I1}-\eqref{I3} in what follows. 


The motion of a free spinning particle in a general translationally accelerated reference frame is considered in the following section and geodetic precession in the Schwarzschild geometry is briefly treated in classical and quantum regimes in sections III and IV, respectively, in order to clarify certain issues that exist in previous papers on this subject---see~\cite{Lee},~ \cite{VR} and the references cited therein. In both cases the spacetime metric has the diagonal form
\begin{equation}\label{metric}
 g_{00} = -\,V^2\,,  \qquad   g_{ij} = W^2\,\delta_{ij}\,.
\end{equation}
In particular, we emphasize that spin precession is not a local phenomenon and there is no basic reason a priori to expect that Einstein's extremely local principle of equivalence would apply in this case. We show explicitly that it does not apply in the ``gravitoelectric" case.  It is curious, however, that it does work for the case of rotation~\cite{M1}. This is due to the fact that for a general noninertial observer, spin does not couple in the same way to an observer's translational acceleration (i.e., 4-acceleration) and the rotation of its spatial frame; in particular, there is no \emph{direct} analog of the spin-rotation coupling in the case of translational acceleration~\cite{BCM}.

In section V, we discuss spin precession in the teleparallel equivalent of general relativity, which involves a tetrad approach to gravitation that is of current interest. Finally, section VI contains a brief discussion of our results.

\section{Spin Precession in a Translationally Accelerated System}\label{trans}

Imagine an observer following an arbitrary accelerated path in a global inertial frame in Minkowski spacetime. The observer carries along its worldline a nonrotating (i.e., Fermi-Walker transported) spatial frame. This spatial frame and the observer's 4-velocity constitute the observer's orthonormal tetrad frame. Based on this tetrad frame, we establish a natural geodesic normal coordinate system in the neighborhood of the worldline of the accelerated observer. This (Fermi) coordinate system $(ct, {\bf x})$ has a metric tensor (\ref{metric}) with~\cite{CM1}
\begin{equation}\label{II1}
 V = 1 + {\frac {\mathbf{a}(t) \cdot \mathbf{x}} {c^2}}\,,\qquad W = 1. 
\end{equation}
The accelerated observer occupies the spatial origin of this coordinate system $(\mathbf{x}=0)$, $t$ is its proper time and the projection of its 4-acceleration vector on its tetrad frame is given by $(0, \mathbf{a})$, where $\mathbf{a}(t)$ is thus the invariant translational acceleration of the observer. The coordinate system is admissible so long as $V \ne 0$. 

Consider now a \emph{free} spinning test particle in this spacetime region. The particle moves with velocity $\mathbf{v}=d\mathbf{x}/dt$ and carries spin vector $S^\mu$; therefore, we can write Eq.~\eqref{I1} in the form
\begin{equation}\label{II2}
 V^2 S^0 = {\frac {\mathbf{v} \cdot \mathbf{S}}{c}}\,, 
\quad  {\frac {dS^i}{dt}} + {\frac {a^i}c} VS^0=0\,,
\end{equation}
since the only nonzero components of the Christoffel symbols are
\begin{equation}\label{II3}
\Gamma^0_{00} = {\frac {\dot{\mathbf{a}} \cdot \mathbf{x}}{c^3V}} \,, 
\quad \Gamma^0_{0i} = {\frac{a_i}{c^2V}}\,, \quad \Gamma^i_{00} = {\frac {Va_i}{c^2}}\,.
\end{equation}

To determine spin precession unambiguously with respect to the accelerated frame of reference, we need to define an orthonormal tetrad frame that is comoving with the spinning particle such that  its spatial frame does not rotate with respect to the background reference axes.  In such a local rest  frame $E^\mu{}_{\hat{\alpha}}$, where $E^\mu{}_{\hat{0}}={\frac 1c}u^\mu$ is the 4-velocity of the spinning particle divided by $c$, the spin vector is purely spatial, namely, $S_{\hat{\alpha}}=S_\mu E^\mu{}_{\hat{\alpha}}=(0, S_{\hat{i}})$. This tetrad frame is derived in Appendix~\ref{appA}, where we write $u^\mu=\Gamma(c, \mathbf{v})$ and $\Gamma^{-1}=\sqrt{V^2-v^2/c^2}$ is the modified Lorentz factor of the particle in this case. We find, using the results of Appendix~\ref{appA}, that
\begin{equation}\label{II4}
S_{\hat{i}} = S_i -\frac{(\Gamma V-1)}{\Gamma V v^2} (\mathbf{v} \cdot \mathbf{S})~v_i\,.
\end{equation}
It is possible to show that in general this locally ``measured" spin vector undergoes precession; in fact, this is demonstrated in Appendix~\ref{appB}, where we derive a general expression for the precession frequency. 

It proves useful at this point to limit our considerations to terms that are at most of order $1/c^2$. In this connection, we note that Eq.~\eqref{II1} yields $\Gamma V-1=v^2/(2c^2)+\dots$, where the dots denote higher order terms that we neglect. Thus Eq.~\eqref{II4} reduces to
\begin{equation}\label{II5}
S_{\hat{i}} = S_i - {\frac{1}{2 c^2}} (\mathbf{v} \cdot \mathbf{S})~v_i + \dots\,.
\end{equation}
Differentiating this equation with respect to time $t$ and using Eq.~\eqref{II2}, we find that up to order $c^{-2}$,
\begin{equation}\label{II6}
{\frac{dS_{\hat{i}}}{dt}} = -\,{\frac{1}{c^2}}(\mathbf{v} \cdot \mathbf{S})a_i - {\frac{1}{2 c^2}}(\frac{d\mathbf{v}}{dt}\cdot\mathbf{S})v_i - {\frac{1}{2 c^2}}(\mathbf{v}\cdot\mathbf{S}){\frac{dv_i}{dt}}\,.
\end{equation}
On the other hand, it follows from the reduced geodesic equation of motion of the free particle that~\cite{CM1} 
\begin{equation}\label{II7}
{\frac{dv_i}{dt}} = {\frac{(\dot{\mathbf{a}}\cdot\mathbf{x} + 2~\mathbf{a}\cdot\mathbf{v})}{c^2V}}~v_i - Va_i\,. 
\end{equation}
Therefore, 
\begin{equation}\label{II8}
{\frac{dv_i}{dt}} = -\,a_i + {\cal O}({\frac{1}{c^2}})\,. 
\end{equation}
Substituting this relation in Eq.~\eqref{II6}, we find that up to order $c^{-2}$
\begin{equation}\label{II9}
{\frac {dS_{\hat{i}}}{dt}} = -\,{\frac{1}{2c^2}}(\mathbf{v}\cdot\mathbf{S})a_i + {\frac {1}{2 c^2}}(\mathbf{a}\cdot\mathbf{S})v_i\,.
\end{equation}
Taking into account the proper time $\tau$ of the spinning particle, Eq.~\eqref{II9} can be written as
\begin{equation}\label{II10}
{\frac {dS_{\hat{i}}}{d\tau}} = \epsilon_{ijk}~{}^{(a)}\Omega_j~ S_{\hat{k}}\,,
\end{equation}
where 
\begin{equation}\label{II11}
{}^{(a)}\boldsymbol{\Omega} = {\frac{\mathbf{a} \times \mathbf{v}}{2c^2}} + \dots
\end{equation}
is the instantaneous precession frequency of the spin relative to the comoving spatial axes that do not rotate with respect to the background frame and the dots represent terms of order higher than $c^{-2}$. 

Several remarks are in order at this point. The precession frequency vanishes if the free particle moves along the direction of acceleration of the fiducial observer. Moreover, up to order $c^{-2}$, it makes no difference if the frequency of precessional motion is referred to the coordinate time $t$ (which is the proper time of the fiducial observer) or the proper time $\tau$ of the free particle. Our result, Eq.~\eqref{II11}, appears to be similar in form to the Thomas precession frequency (see Appendix~\ref{appC}), but this analogy is at best misleading, since in Thomas precession $\mathbf{a}$ would be the acceleration of the spinning particle. Note that if in Eq.~\eqref{II11} we replace $\mathbf{a}$ by $-d\mathbf{v}/dt$ in accordance with Eq.~\eqref{II8}, we get minus the expected result from an invalid application of the Thomas precession formula~\cite{Lee}. The fact is that accelerated motion is absolute in relativity theory and the two situations are not directly related. Finally, Eq.~\eqref{II11} agrees completely with the result of Hehl and Ni~\cite{HN}; that is, in the quantum domain, the precession of the spin would be naturally attributed to the Hamiltonian
\begin{equation}\label{II12}
{}^{(a)}H_{spin} = {\frac{1}{2mc^2}}(\mathbf{a} \times \mathbf{p}) \cdot \mathbf{S}\,,
\end{equation}
where $\mathbf{p}=m\mathbf{v}$ to lowest order. For a Dirac particle of spin $\frac{\hbar}{2}$, this Hamiltonian agrees with the ``new inertial spin-orbit coupling" first elucidated in Ref.~\cite{HN} for  Dirac particles in accelerated frames of reference via Foldy-Wouthuysen transformations.  In connection with this concordance between classical and quantum results, we emphasize that in Ref.~\cite{HN}, the \emph{same} basic background reference frame has been employed as in our classical treatment in this section. 

An electron carries both electric charge and intrinsic spin $\frac{\hbar}{2}$, and the study of spin-dependent electron transport phenomena has led to the emerging field of spintronics. It is therefore of interest to study spin currents in noninertial frames~\cite{GP}. Some of the mechanical effects of  rotation on spin currents as well as on magnetic resonance phenomena have been the subjects of recent investigations~\cite{S1, S2, S3, MR1, MR2, MR3}.  In this connection, we mention that the results elucidated in this paper may also be of observational interest in the future.  

We next turn to the coupling of spin with the gravitoelectric field.

\section{Spin Precession in the Schwarzschild Field}\label{schwarz}

It turns out that for the considerations of this section, namely, the calculation of spin precession up to order $c^{-2}$ in Schwarzschild spacetime, the linear approximation to general relativity is sufficient~\cite{M1}. Thus, to linear order, the nonzero components of the exterior Schwarzschild metric in isotropic coordinates $(ct, \mathbf{x})$ are given by Eq.~(\ref{metric}) with
\begin{equation}\label{III1}
V = 1 - \phi \,,  \qquad  W = 1 + \phi\,,
\end{equation}
where $\phi=GM/(c^2r)$, $\phi \ll 1$, $M$ is the mass of the spherical source and $r=|\mathbf{x}|$. It is straightforward to show that Eq.~\eqref{I1} reduces to order $c^{-2}$ to $S^0=(\mathbf{v} \cdot \mathbf{S})/c$ and
\begin{equation}\label{III2}
{\frac{dS^i}{dt}} + \phi_{,j}S^j v^i + \phi_{,j}v^j S^i - 2(\mathbf{v}\cdot\mathbf{S})\phi_{,i} =0\,.
\end{equation}
Here $\phi_{,i} = \partial_i\,\phi=c^{-2}g_i$, where $\mathbf{g}$ is the Newtonian acceleration of gravity and Eq.~\eqref{I2} for the motion of the spinning test particle reduces in this case to 
\begin{equation}\label{III3}
{\frac{dv_{i}}{dt}} = g_i + {\cal O}({\frac{1}{c^2}})\,. 
\end{equation}

We are interested in the motion of the spin relative to an orthonormal comoving frame $E^\mu{}_{\hat{\alpha}}$, given in Appendix~\ref{appA}, that consists of the temporal axis $u^\mu$ and three spatial axes that are boosted without any rotation with respect to the background spatial axes of the spacetime. The advantage of this procedure is that it exhibits the pure motion of the spatial spin vector, as $S_{\hat{0}}=0$ by definition. It follows from the results of Appendix~\ref{appA} that $S_{\hat{\alpha}} =S_\mu E^\mu{}_{\hat{\alpha}}=(0, S_{\hat{i}})$, where up to order $c^{-2}$
\begin{equation}\label{III4}
S_{\hat{i}}= (1 + \phi)S_i - {\frac{1}{2 c^2}}(\mathbf{v}\cdot\mathbf{S})~v_i\,.
\end{equation}
Differentiating this expression with respect to time $t$, using Eqs.~\eqref{III2} and~\eqref{III3} and expressing the end result in terms of the proper time $\tau$ of the spinning particle, we find that \begin{equation}\label{III5}
{\frac{dS_{\hat{i}}}{d\tau}} = \epsilon_{ijk}~{}^{(g)}\Omega_j~ S_{\hat{k}}\,,
\end{equation}
where, up to order $c^{-2}$,
\begin{equation}\label{III6}
{}^{(g)}\boldsymbol{\Omega} = -\,\frac{3(\mathbf{g} \times \mathbf{v})}{2c^2}
\end{equation}
is the well-known geodetic (i.e., de Sitter-Fokker) precession frequency~\cite{SRCB}. It has recently been directly measured in Earth orbit via the GP-B experiment~\cite{EV}. 

It is important to note that replacing $-\mathbf{g}$ by $\mathbf{a}$ does not turn Eq.~\eqref{III6} into Eq.~\eqref{II11}; in fact, the gravitational effect is three times larger in magnitude than would be  expected from a naive application of Einstein's principle of equivalence. 

Einstein's heuristic principle of equivalence is the cornerstone of general relativity theory~\cite{AE}, upon which our calculations are based. On the other hand, a simple and direct application of the principle involves replacing minus the acceleration of gravity by the translational acceleration of the frame in Minkowski spacetime. We know, for instance, that to order $c^{-2}$ this procedure would predict only half of the bending of light in the exterior Schwarzschild geometry. Referring to Eqs.~\eqref{III1} and~\eqref{II1}, one might interpret  the general relativity result for light bending as being half  due to temporal curvature and the other half due to spatial curvature, since space is not flat in Eq.~\eqref{III1} in contrast to Eq.~\eqref{II1}. In a similar way, we can compare and contrast our calculation of the geodetic precession with that of the previous section and thereby characterize the influence of the spatial part of the metric on the calculation of the precession rate. In this way, the origin of the factor of three can be identified: One part of this factor is due to temporal curvature---just as in the case of translational acceleration---and the other two parts are due to spatial curvature. 

In the quantum domain, the analog of Eq.~\eqref{II12} in this case would be a gravitational spin-orbit coupling given by the Hamiltonian
\begin{equation}\label{III7}
{}^{(g)}H_{spin} = -\,{\frac{3}{2mc^2}}(\mathbf{g}\times\mathbf{p})\cdot\mathbf{S}\,.
\end{equation}
The existence of such a coupling for a Dirac particle is demonstrated in section~\ref{dirac}.

It is important to point out another situation where the motion of the spin vector can be unambiguously studied. We recall that up to order $c^{-2}$, it is sufficient to take due account of only the \emph{Newtonian} orbit of the particle. Neglecting scattering orbits of the spinning particle, we note that the bounded orbits reduce in our approximation scheme to planar Keplerian ellipses that perform a ``fast" motion of period $T_K$. On the other hand, any motion of the spatial spin vector would be ``slow", with a long period proportional to $c^2$. To bring out this spin motion clearly, it is therefore possible to average over the fast motion. The averaging procedure is defined as usual by 
\begin{equation}\label{III8}
\langle f \rangle = {\frac{1}{T_K}} \int\limits_0^{T_K} f dt\,.
\end{equation}
Let us assume, without any loss in generality, that the background axes are so oriented that the Keplerian ellipse corresponding to the Newtonian motion of the particle is given by $x^1=\rho \cos{\varphi}$, $x^2=\rho \sin{\varphi}$ and $x^3=0$, where
\begin{equation}\label{III9}
\rho = {\frac{A(1-e^2)}{1+e\cos{\varphi}}}\,,\qquad {\frac{d\varphi}{dt}} = {\frac{\ell}{\rho^2}}\,.
\end{equation}
Here $A$ and $e$ are respectively the semimajor axis and the eccentricity of the ellipse and the Keplerian period is given by $T_K=2\pi A (GM/A)^{-1/2}$. Moreover, $\boldsymbol{\ell}$ is the vector of  specific orbital angular momentum of the Newtonian orbit and points along the $x^3$ 
axis, while $\ell=|\boldsymbol{\ell}|=[GMA(1-e^2)]^{1/2}$. In terms of the azimuthal angle $\varphi$, the averaging takes the form
\begin{equation}\label{III10}
\langle f \rangle = {\frac{(1-e^2)^{3/2}}{2\pi}} \int\limits_0^{2\pi} {\frac{f(\varphi)}{(1+e\cos{\varphi})^2}}~d\varphi \,.
\end{equation}

It follows from our averaging procedure that $\langle \mathbf{v} \rangle=0$ and $\langle d\phi/dt \rangle=0$ for any closed Newtonian orbit and hence from the equations of motion for the spin vector we conclude that \emph{on average} $S^0=0$. Moreover, in Eq.~\eqref{III2}, we find upon averaging that
\begin{equation}\label{III11}
\left \langle {\frac{x^iv^j}{\rho^3}}\right \rangle = {\frac{1}{2}}\left \langle {\frac{1}{\rho^3}}\right \rangle \epsilon_{ijk} \ell^k\,,
\end{equation}
where
\begin{equation}\label{III12}
\left \langle {\frac{1}{\rho^3}} \right \rangle = {\frac{1}{A^3(1-e^2)^{3/2}}}\,.
\end{equation}
Therefore, Eq.~\eqref{III2} implies that on the average 
\begin{equation}\label{III13}
{\frac{dS_i}{dt}} = \epsilon_{ijk}~ \langle ^{(g)}\Omega_j \rangle~ S_{k}\,,
\end{equation}
where
\begin{equation}\label{III14}
\langle ^{(g)}\boldsymbol{\Omega}\rangle = \left \langle -\,{\frac{3}{2c^2}}(\mathbf{g} \times \mathbf{v})\right \rangle = {\frac{3}{2c^2}}~\frac{GM \boldsymbol{\ell}}{A^3(1-e^2)^{3/2}}\,.
\end{equation}
We note that for a circular orbit, this average precession frequency coincides with the instantaneous result given in Eq.~\eqref{III6}.

The formal results given in Eqs.~\eqref{III13} and~\eqref{III14} for the coordinate components of spin $S^\mu$ acquire physical significance once they are referred to the tetrad frame of a suitable family of observers. Imagine, for instance, the class of \emph{fundamental observers at rest} all along the orbit of the spinning particle. We assume that these observers refer their spin measurements to their natural tetrad frames: each such observer has a temporal axis $(1+\phi)\delta^\mu_0$ and spatial axes $(1-\phi)\delta^\mu_i$ in our linear approximation scheme ($\phi \ll 1$). A detailed investigation reveals that, up to order $c^{-2}$, we recover Eq.~\eqref{III14} upon averaging. 

The manifestation of geodetic precession in the quantum domain is the subject of the next section.

\section{Dirac Particle in Schwarzschild Spacetime}\label{dirac}

The covariant Dirac equation for spin-1/2 particles in a
gravitational (or inertial) field is given by
\begin{equation}
(i\hbar\gamma^{\hat \alpha} D_{\hat \alpha} - mc)\Psi=0\,.
\label{Dirac0}
\end{equation}
The flat Dirac matrices $\gamma^{\hat \alpha}$ are defined in local Lorentz
(tetrad) frames $e_\mu{}^{\hat \alpha}$. For the class of diagonal metrics (\ref{metric}),
we choose $e_\mu{}^{\hat 0} = V\delta^0_\mu$ and $e_\mu{}^{\hat i} = W\delta^i_\mu$. 
The spinor covariant derivatives are given by
\begin{equation}
D_{\hat \alpha} = e^\mu{}_{\hat \alpha}D_\mu\,,\qquad D_\mu = \partial _\mu 
+ {\frac i4}\sigma^{{\hat \alpha}{\hat \beta}}\Gamma_{\mu\,{\hat \alpha}{\hat \beta}}\,.\label{eqin2}
\end{equation}
Here $\Gamma_\mu{}^{{\hat \alpha}{\hat \beta}} = - \Gamma_\mu{}^{{\hat \beta}{\hat \alpha}}$ are
the Lorentz connection coefficients and $\sigma^{{\hat \alpha}{\hat \beta}} = \frac
i2\left(\gamma^{\hat \alpha} \gamma^{\hat \beta} - \gamma^{\hat \beta}\gamma^{\hat \alpha}\right)$. 

We can derive the Dirac equation from the action
\begin{equation}
I = \int\,d^4x\,{\cal L}\,,\qquad {\cal L} = \sqrt{-g}\,L\label{action}
\end{equation}
with the Lagrangian
\begin{equation}\label{LD}
L = {\frac {i\hbar}{2}}\left(\overline{\Psi}\gamma^{\hat \alpha}
D_{\hat \alpha}\Psi - D_{\hat \alpha}\overline{\Psi}\gamma^{\hat \alpha}\Psi\right) -
mc\,\overline{\Psi}\Psi\,.
\end{equation}
The naive Hamiltonian for the Schr\"odinger form of the Dirac equation, derived from  
action (\ref{action}), is not Hermitian. In order to solve this hermiticity problem, 
we need to redefine the wave function as
\begin{equation}
\psi = \left(\sqrt{-g}~e^0{}_{\hat 0}\right)^{\frac
12}\,\Psi = W^{\frac 32}\,\Psi.\label{newpsi}
\end{equation}
As a result, the variation of the action with respect to the rescaled wave function yields the Dirac equation in Schr\"odinger form $i\hbar\frac{\partial \psi} {\partial t}= 
{\cal H}\psi$. The corresponding {\it Hermitian} Hamiltonian~\cite{ynoprl,st1,st2}, 
for metric (\ref{metric}), can be expressed as
\begin{equation}\label{Hamiltonian1}
{\cal H} = \beta mc^2V + {\frac c 2}\left[(\boldsymbol{\alpha}
\cdot \mathbf{p}){\cal F} + {\cal F}(\boldsymbol{\alpha}\cdot \mathbf{p})\right].
\end{equation}
Here ${\cal F} := V/W$ and $\mathbf{p}$ is the momentum operator, $\mathbf{p} = -\,i\hbar\boldsymbol{\nabla}$. As usual,
we denote $\beta = \gamma^{\hat 0}$ and $\alpha^i = \beta\gamma^{\hat i}$. 

The physical content of the theory is revealed in the Foldy-Wouthuysen representation.
After performing the Foldy-Wouthuysen transformation, the Hamiltonian is recast in the 
semiclassical approximation~\cite{ost1,ost2} into the form
\begin{equation}
{\cal H}_{FW} = \beta\left(\epsilon  + \frac\hbar2\boldsymbol{\Sigma}\cdot \boldsymbol{\Omega}\right).\label{HFW}
\end{equation}
Here $\boldsymbol{\Sigma}$ is the spin operator, $\epsilon=\sqrt{m^2c^4V^2+c^2 p^2{\cal F}^2}$, $p=|\mathbf{p}|$ and
\begin{equation}
\boldsymbol{\Omega} = {\frac {\mathbf{p}\times\mbox{\boldmath ${\cal A}$}}{mc^2}}\,,\qquad 
\mbox{\boldmath ${\cal A}$} = {\frac{mc^4{\cal F}}{\epsilon}}\left({\frac {mc^2{\cal F}}
{\epsilon + mc^2V}}\boldsymbol{\nabla}V - \boldsymbol{\nabla}{\cal F}\right).\label{omega}
\end{equation}
Keeping only the leading terms in $c^{-2}$ expansion, for the Schwarzschild metric 
we have $\epsilon= mc^2V$, $V = 1 - \phi$, ${\cal F} = 1 - 2\phi$ and we find 
$\mbox{\boldmath ${\cal A}$} = 3\mathbf{g}/2$, in perfect agreement with the classical 
result (\ref{III6}). 

We can straightforwardly compare this result with the dynamics of quantum spin in the
noninertial reference frame of section II. Substituting Eq.~\eqref{II1} in Eq.~\eqref{omega}, we obtain
$\mbox{\boldmath ${\cal A}$} = -\,\mathbf{a}/2$ in complete agreement with Ref.~\cite{HN}. 

Many aspects of the behavior of Dirac particles in gravitational and inertial fields have been extensively studied in the past---see, for instance, Ref.~\cite{ost2} for further references and discussion. The purpose of the brief account presented here has been to demonstrate explicitly the complete consistency of the classical and quantum approaches to the problem of spin precession. This follows from the general correspondence principle and the circumstance that, in either approach, the \emph{same} orthonormal tetrad frame is employed for the fundamental background observers~\cite{LR}.  To illustrate this latter point, let us employ a different spatial frame for the fundamental observers at rest in the Schwarzschild field. In the linear approximation, the standard form of the Schwarzschild metric can be expressed in Cartesian coordinates~\cite{VR} as follows: 
\begin{equation}
g_{00} = -\,V^2,\qquad g_{0i} = 0,\qquad 
g_{ij} = {\cal W}^{\hat k}{}_i{\cal W}^{\hat l}{}_j\delta_{{\hat k}{\hat l}}.\label{metricS}
\end{equation}
Here $V = 1 - \phi$, ${\cal W}^{\hat k}{}_i = \delta^k_i + \phi\,x^kx_i/r^2$, the spatial indices are raised and lowered using the Euclidean 3-metric $\delta_{ij}$ and $r^2 = \delta_{ij}x^ix^j$. We note that metric (\ref{metric}) belongs to the general family (\ref{metricS}), since we can choose ${\cal W}^{\hat k}{}_i = W\delta^k_i$. 

The tetrad coframe $e_\mu{}^{\hat \alpha}$ of the background fundamental observers at rest would then have a temporal axis given as before by $e_\mu{}^{\hat 0} = V\delta^0_\mu$ and spatial axes that are now $e_\mu{}^{\hat i} = {\cal W}^{\hat i}{}_j\delta^j_\mu$. This spatial frame coincides with the old one at spatial infinity. Starting from this new background frame, we can construct a new boosted tetrad frame along the path of a free spinning test particle as in section III and Appendix A, and show explicitly that to order $c^{-2}$ in the classical regime, particle spin precesses with the geodetic precession frequency precisely as in Eqs.~\eqref{III5} and~\eqref{III6}. This means that to order $c^{-2}$, no dynamic rotation is involved here; indeed, the new and old spatial frames are Fermi-Walker transported along the worldlines of the fundamental static observers. 

In the quantum approach, we therefore expect to recover the same result as before; in fact, this also follows from the work of Varj\'u and Ryder~\cite{VR} once a computational error is corrected. The Hermitian Dirac Hamiltonian in the new Schwarzschild coordinates can now be expressed as 
\begin{equation}\label{Hamiltonian2}
{\cal H} = \beta mc^2V + {\frac c 2}\left[p_i~{\cal F}^i{}_k~\alpha^k + \alpha^k~{\cal F}^i{}_k~p_i\right].
\end{equation}
Here ${\cal F}^i{}_k  := V {\cal W}^i{}_{\hat k}$, where the matrix ${\cal W}^i{}_{\hat k}$ is defined to be the inverse of the matrix ${\cal W}^{\hat k}{}_i$. For ${\cal W}^{\hat k}{}_i = W\delta^k_i$, we obviously recover Eq.~(\ref{Hamiltonian1}). Performing the Foldy-Wouthuysen transformation, we obtain the semiclassical Hamiltonian (\ref{HFW}) such that the first term is now generalized to $\epsilon=\sqrt{m^2c^4V^2+c^2\delta^{kl}{\cal F}^i{}_k{\cal F}^j{}_lp_ip_j}$; moreover, the components of the precession angular velocity are now given by
\begin{equation}
\Omega^i = {\frac {c^2}{\epsilon}}\,{\cal F}^l{}_n\,p_l\left(-\,\epsilon^{{\hat i}{\hat j}{\hat k}}\,VC_{{\hat j}{\hat k}}{}^{\hat n} + {\frac {\epsilon}{\epsilon + mc^2V}}\epsilon^{{\hat i}{\hat j}{\hat n}}\,{\cal W}^l{}_{\hat j}\partial_lV\right).\label{OmegaR}
\end{equation}
Here the anholonomity object is defined as usual by $C_{{\hat j}{\hat k}}{}^{\hat n} = {\cal W}^i{}_{\hat j}{\cal W}^l{}_{\hat k}\partial_{[i}{\cal W}^{\hat n}{}_{l]}$. One can verify that for the diagonal case ${\cal W}^{\hat k}{}_i = W\delta^k_i$, the general formula (\ref{OmegaR}) reduces to the simplified result (\ref{omega}). Keeping again only the leading terms in $c^{-2}$ expansion, for  metric (\ref{metricS}) we have $\epsilon= mc^2V$, $V = 1 - \phi$ and $C_{{\hat j}{\hat k}}{}^{\hat n} = -\,\delta^n{}_{[j}~g_{k]}/c^2$; therefore,  we find $\boldsymbol{\Omega} = 3\mathbf{p}\times\mathbf{g}/(2mc^2)$, in perfect agreement with the corresponding classical result. On the other hand, the numerical coefficient of $\boldsymbol{\Omega}$ deduced from Ref.~\cite{VR} would be unity instead of our $3/2$.

Our new computation thus confirms that, contrary to previous reports~\cite{Lee,VR}, the classical and quantum pictures are completely consistent irrespectively of the local coordinates used---namely, the isotropic form of metric (\ref{metric}) linearized via Eq.~(\ref{III1}), or the linearized Schwarzschild metric  in Cartesian coordinates (\ref{metricS}). We suspect that the origin of the erroneous numerical coefficient in front of gravitational spin-orbit coupling term in the final Hamiltonian of Ref.~\cite{VR} is buried in the details of the evaluation of the commutators in the Foldy-Wouthuysen transformation.

\section{Spin Precession in GR$_{||}$}\label{Tele}

Teleparallel gravity (``GR$_{||}$"), also known as the teleparallel equivalent of general relativity, is a viable alternative tetrad theory constructed within the framework of the gauge approach to the gravitational interaction---see~\cite{reader,AP,Maluf} for recent comprehensive treatments of this subject. GR$_{||}$ is a gauge theory of the group of spacetime translations. In this model,  the gravitational field is described via the coframe (or tetrad) $e_{\mu}{}^{\hat \alpha}$, in terms of which the Weitzenb\"ock connection is constructed that has vanishing curvature (``distant parallelism"), but is instead characterized by the nontrivial torsion tensor
\begin{equation}\label{torsion}
T_{\mu\nu}{}^\alpha = e^{\alpha}{}_{\hat \beta}\left(
\partial_\mu e_{\nu}{}^{\hat \beta} - \partial_\nu e_{\mu}{}^{\hat \beta}\right).
\end{equation}
It is important to stress that for matter source without intrinsic spin, teleparallel gravity is essentially indistinguishable from Einstein's general relativity theory (``GR")~\cite{HS}. Furthermore, the coupling of the spinning matter to the coframe in GR$_{||}$ is in general \emph{ inconsistent}~\cite{OP1,OP2}. More exactly, one can construct a consistent coupling if one assumes that the Dirac spinor fermion field interacts with the gravitational field by means of the usual \emph{Riemannian} connection in the framework of standard minimal coupling scheme (\ref{eqin2}). In this case, the precession results obtained for GR will be valid {\it mutatis mutandis} in GR$_{||}$ as well.

Suppose, on the other hand, that instead of the Riemannian connection, we use the Weitzenb\"ock connection, which would be more in keeping with the spirit of the gauge-theoretic approach; then, the Dirac field Lagrangian is only invariant under \emph{global} Lorentz transformations. This circumstance is in sharp contrast to the invariance of the gravitational Lagrangian under \emph{local} Lorentz transformations of the tetrad fields. In this case, the dynamics of particles with spin will be \emph{different} in GR than in GR$_{||}$~\cite{HS,Nitsch,Rumpf,Zhang}; indeed, making use of the formalism developed in Sec.~\ref{dirac}, we can derive the Foldy-Wouthuysen Hamiltonian for the Dirac particle in this teleparallel model. Using the Weitzenb\"ock connection, the influence of the torsion tensor (\ref{torsion}) shows up in a new contribution to the precession frequency. In fact, we find that the precession frequency $\boldsymbol{\Omega}$ in Eq. (\ref{omega}) is replaced by $\boldsymbol{\Omega} + \Delta\boldsymbol{\Omega}$, where
\begin{equation}\label{omegaT}
\Delta\boldsymbol{\Omega} = {\frac {3cV}{2}}\left(\boldsymbol{\check{T}}
- {\frac {c{\cal F}}{\epsilon}}\,\mathbf{p}\,\check{T}^{\hat 0}\right).
\end{equation}
Here $\check{T}^{\hat 0}$ and $\check{T}^{\hat i} (=\boldsymbol{\check{T}})$ are the components of the axial torsion pseudovector
\begin{equation}
\check{T}^{\hat\alpha} = {\frac 16}\,\epsilon^{{\hat\mu}{\hat\nu}{\hat\lambda}
{\hat\alpha}}\,T_{{\hat\mu}{\hat\nu}{\hat\lambda}}.\label{Tcheck}
\end{equation}
Our result (\ref{omegaT}) agrees with previous treatments of this issue~\cite{HS,Nitsch,Rumpf,Zhang}. When we specialize to the specific cases discussed in the present paper (i.e., the static gravitational field or the translational acceleration), the axial part of the Weitzenb\"ock torsion (\ref{Tcheck}) vanishes. Our conclusions in this paper thus remain valid for teleparallel gravity precisely in the same form as for Einstein's general relativity. However, the torsion components in $\boldsymbol{\check{T}}$ are in general nontrivial for stationary gravitational field configurations, such as the Kerr spacetime, as well as for rotating systems in Minkowski spacetime. The corresponding additional contribution to the precession frequency, given in the linear weak-field approximation by~$\Delta \boldsymbol{\Omega} = {\frac {3c}{2}}\,\boldsymbol{\check{T}}$, should then be combined with similar terms arising from the total spin $\mathbf{J}$ of the gravitational source or the angular velocity $\boldsymbol{\omega}$ of the rotating reference frame \cite{ost2}.

\section{Discussion}\label{discuss}

The main purpose of this paper has been to elucidate the classical and quantum aspects of spin-orbit coupling in inertial and gravitational fields. There is no analog of the coupling of spin to rotation and gravitomagnetic fields for translational acceleration and gravitoelectric fields and this has been a source of some confusion. In particular, we have addressed the problems raised in Refs.~\cite{Lee} and~\cite{VR}, and have clarified the physics of spin precession in GR and GR$_{||}$, the teleparallel equivalent of general relativity. 

\begin{acknowledgments}
We are grateful to Friedrich Hehl and Lewis Ryder for valuable discussions. 
\end{acknowledgments}

\appendix{}
\section{Boosted Tetrads}\label{appA}

Consider the fundamental observers at rest in an inertial frame with coordinates $(ct, \mathbf{x})$ in Minkowski spacetime. An inertial observer moves with constant velocity $\mathbf{v}=v \hat{\mathbf{t}}$ with respect to the fundamental observers. Here $\hat{\mathbf{t}}$ is the tangential unit vector. The Lorentz transformation to the rest frame of the moving observer $(ct', \mathbf{x'})$ involving a pure boost with no rotation is given by
\begin{eqnarray}\label{A1}
t &=& \gamma(t'+\mathbf{v} \cdot \mathbf{x'})\,,\\
\mathbf{x} &=& \mathbf{x'}+(\gamma -1)(\mathbf{x'} \cdot \hat{\mathbf{t}})\hat{\mathbf{t}} +\gamma \mathbf{v} t \,,\label{A2}
\end{eqnarray}
where $\gamma$ is the Lorentz factor. In the $(ct', \mathbf{x'})$ frame, the tetrad of the inertial observers at rest is given by $h'^{\mu}{}_{\hat{\alpha}}=\delta^\mu_\alpha$\,. Transforming the local tetrad frame of the moving observer to the $(ct, \mathbf{x})$ system via the Lorentz boost matrix $\Lambda_b$ that can be simply deduced from Eqs.~\eqref{A1} and~\eqref{A2}, we find
\begin{eqnarray}\label{A3}
h^{\mu}{}_{\hat{0}} &=& \gamma \left(1, {\frac {\mathbf{v}}{c}}\right)\,,\\
h^{\mu}{}_{\hat{i}} &=& \delta^\mu_i + v_i\!\left({\frac \gamma c}, \frac{(\gamma - 1)}{v^2} \mathbf{v}\right)\,.\label{A4}
\end{eqnarray}
This is the tetrad of the boosted observer with respect to the fundamental observers at rest in the $(ct, \mathbf{x})$ system. We are interested in the generalization of this result to arbitrary spacetimes.

Let us first assume that the metric of the background spacetime is not $\eta_{\mu \nu}$, but instead $g_{\mu \nu}$, where the difference between them is  in the purely temporal component, namely, $g_{00}=-\,V^2(t, \mathbf{x})$, as in section II. We are interested in the local comoving tetrad frame $E^\mu{}_{\hat{\alpha}}$. It is straightforward to conclude via inspection that the generalization of Eqs.~\eqref{A3} and~\eqref{A4} is given in this case by
\begin{eqnarray}\label{A5}
E^{\mu}{}_{\hat{0}} &=& \Gamma \left(1, {\frac {\mathbf{v}}{c}}\right)\,,\\
E^{\mu}{}_{\hat{i}} &=& \delta^\mu_i + v_i \!\left({\frac{\Gamma}{Vc}}, \frac{(\Gamma V - 1)}{v^2} \mathbf{v}\right)\,,\label{A6}
\end{eqnarray}
where $\Gamma^{-1}=\sqrt{V^2-v^2/c^2}$. That is, Eqs.~\eqref{A5} and~\eqref{A6} are such that the temporal axis is the observer's 4-velocity and
\begin{equation}\label{A7}
g_{\mu \nu} E^\mu{}_{\hat{\alpha}}E^\nu{}_{\hat{\beta}}=\eta_{{\hat{\alpha}}{\hat{\beta}}}\,;
\end{equation}
moreover, they agree with Eqs.~\eqref{A3} and~\eqref{A4} for $V=1$ and reduce for $\mathbf{v}=0$ to the natural tetrad frame of the fundamental observers at rest.

As our second example, we consider the generalization of Eqs.~\eqref{A3} and~\eqref{A4} to the case of  metric (\ref{metric}), where $V>0$ and $W>0$ are positive functions of spacetime coordinates. An important example is the isotropic form of the exterior Schwarzschild metric with  
\begin{equation}\label{A8}
V = \left(\frac{1-\phi/2}{1+\phi /2}\right)\,,\qquad W = \left(1+\frac{\phi}{2}\right)^2\,,
\end{equation}
where $\phi=GM/(c^2 |\mathbf{x}|)$ and $M$ is the mass of the Schwarzschild source. The metric in this general case is formally related to the previous metric via a conformal transformation. It is an immediate consequence of Eq.~\eqref{A7} that if $g_{\mu \nu} \mapsto W^2 g_{\mu \nu}$, then the corresponding tetrad components must all be multiplied by $W^{-1}$, which is the inverse of the conformal factor $W$. It thus follows from Eqs.~\eqref{A5} and~\eqref{A6}, via $V\mapsto {\cal F} = V/W$, that in this case the relevant tetrads are
\begin{eqnarray}\label{A9}
E^{\mu}{}_{\hat{0}} &=& \frac{\Gamma}{W} \left(1, \mathbf{v}/c\right)\,,\\
E^{\mu}{}_{\hat{i}} &=&\frac{1}{W} \delta^\mu_i + v_i\!\left({\frac{\Gamma}{cV}}, 
\frac{(\Gamma V - W)}{W^2v^2} \mathbf{v}\right)\,,\label{A10}
\end{eqnarray}
where $\Gamma^{-1}=\sqrt{{\cal F}^2 - v^2/c^2}$.

The projection of particle's spin 4-vector on this tetrad is given by  $(0, S_{\hat{i}})$, since $E^{\mu}{}_{\hat{0}}={\frac 1c}u^\mu$ by construction and 
\begin{equation}\label{A11}
V^2S^0 = W^2{\frac {(\mathbf{v} \cdot \mathbf{S})} c}\,.
\end{equation}
Moreover, 
\begin{equation}\label{A12}
S_{\hat{i}}= W\left[S_i -\frac{(\Gamma V - W)}{\Gamma Vv^2} (\mathbf{v} \cdot \mathbf{S}) ~ v_i\right]\,.
\end{equation}
For the Schwarzschild metric with $\phi \ll 1$, keeping terms linear in $\phi$ as well as up to order $c^{-2}$, we have $V = 1 - \phi$, $W = 1 + \phi$, ${\cal F} = 1 - 2\phi$ and $W(\Gamma V - W)/(\Gamma Vv^2)=v^2/(2c^2)$.

For more complicated metrics, the general approach consists of first identifying the natural tetrad frame of the fundamental static observers in spacetime. For instance, for an asymptotically flat geometry, we may choose the frame of the fundamental observers to agree with that of the inertial observers at spatial infinity. Next, at an arbitrary event along the worldline of the particle, we project the particle's 4-velocity $u^\mu$ on the local tetrad of the fundamental observer to get $\gamma_b(c, \mathbf{v}_b)$, where $\gamma_b$ is the Lorentz factor corresponding to $\mathbf{v}_b$, which is the boost velocity that we need to employ in the Lorentz matrix $\Lambda_b$ in order to boost the local tetrad of the fundamental observer to the instantaneous comoving tetrad of the particle. To implement this procedure in the two simple examples above, we need $\mathbf{v}_b=V^{-1} \mathbf{v}$ in the first case and $\mathbf{v}_b=(W/V) \mathbf{v}$ in the second case.

\section{Precession Frequency}\label{appB}

Consider an arbitrary observer following a timelike path in spacetime. The observer carries along its path a local \emph{orthonormal} tetrad frame $\lambda^\mu{}_{\hat{\alpha}}$ such that $\lambda^\mu{}_{\hat{0}}=u^{\mu}/c$ is the temporal axis and $\lambda^\mu{}_{\hat{i}}$, $i=1,2,3,$ are the spatial axes of its local reference frame. The moving frame field in general satisfies
\begin{equation}\label{B1}
{\frac{D\lambda^\mu{}_{\hat{\alpha}}}{d\tau}} = \Phi_{\hat{\alpha}}{}^{\hat{\beta}}~\lambda^\mu{}_{\hat{\beta}}\,,
\end{equation}
where $\tau$ is the observer's proper time and $\Phi_{\hat{\alpha} \hat{\beta}}=-\Phi_{\hat{\beta} \hat{\alpha}}$ is its \emph{antisymmetric acceleration tensor}. This tensor has ``electric" components $\Phi_{\hat{0} \hat{i}}$ given by the components of  4-acceleration $a^{\mu}$ relative to the spatial axes divided by $c$, while its ``magnetic" components $\Phi_{\hat{i} \hat{j}}$ define the angular velocity of rotation of the spatial frame relative to a local nonrotating (i.e., Fermi-Walker transported) frame. The latter can be defined, for instance, by means of ideal free test gyro directions~\cite{M1}. 

Let us next imagine that the observer is comoving with a free spinning test particle along the path given by Eq.~\eqref{I2}. For the spin motion, Fermi-Walker transport reduces to parallel transport, as in Eq.~\eqref{I1}, since in our approximation scheme second-order spin effects are neglected~\cite{CMP}. That is, $u_\mu S^\mu=0$ implies that Fermi-Walker transport reduces to Fermi transport---see Eq.~\eqref{C2} of the next appendix---where, because of Eq.~\eqref{I2}, $a_\nu S^\nu$ is of second order in spin and can therefore be neglected. We then replace $\lambda^\mu{}_{\hat{\alpha}}$ in Eq.~\eqref{B1} with $E^\mu{}_{\hat{\alpha}}$ given in Appendix A and note that the spatial components of the acceleration tensor are now given by
\begin{equation}\label{B2}
\Phi_{\hat{i} \hat{j}}= -\,\epsilon_{ijk} \Omega_k\,,
\end{equation}
where $\boldsymbol{\Omega}$ is the desired frequency of rotation of the particle spin relative to the spatial frame of the boosted tetrad $E^{\mu}{}_{\hat{\alpha}}$. Using Eq.~\eqref{I1}, the covariant derivative of $S_{\hat{i}}=S_\mu E^{\mu}{}_{\hat{i}}$ along the path can be calculated via Eqs.~\eqref{B1} and~\eqref{B2} and the result is
\begin{equation}\label{B3}
{\frac{dS_{\hat{i}}}{d\tau}} = \epsilon_{ijk}~\Omega_j~ S_{\hat{k}}\,.
\end{equation}
From 
\begin{equation}\label{B4}
\Phi_{\hat{i} \hat{j}} = -\,E^{\mu}{}_{\hat{i}}~{\frac {DE_{\mu{}{\hat{j}}}} {d\tau}}\,,
\end{equation}
and Eq.~\eqref{B2}, we find the general expression for the precession frequency, namely, 
\begin{equation}\label{B5}
\Omega_i = {\frac{1}{2}} \epsilon_{ijk}~E^{\mu}{}_{\hat{j}}\left({\frac {dE_{\mu{}\hat{k}}} {d\tau}} - u^\lambda\Gamma_{\lambda \mu}^\nu E_{\nu{}\hat{k}}\right).
\end{equation}
In computing the right-hand side of this equation, the particle's equation of motion~\eqref{I2} must be employed. In view of Eq.~\eqref{B3}, we may simply use the geodesic equation instead of Eq.~\eqref{I2}, as second-order spin effects are neglected in accordance with our general approximation scheme~\cite{CMP}.

\section{Thomas Precession}\label{appC}

Imagine a classical point particle following an arbitrary accelerated path $\mathbf{x}(t)$ in the background global inertial frame and carrying a torque-free ``intrinsic" spin vector $\mathbf{S}$. As is well known, according to the fundamental static inertial observers, the spin vector undergoes Thomas precession with frequency~\cite{Th1, Th2, Jack}
\begin{equation}\label{C1}
\boldsymbol{\omega}_{Thomas} = {\frac{\gamma^2}{\gamma + 1}}~{\frac{\mathbf{a}\times\mathbf{v}}{c^2}}\,,
\end{equation}
where $\mathbf{v}=d\mathbf{x}/dt$ and $\mathbf{a}=d\mathbf{v}/dt$ are the velocity and acceleration vectors of the particle, respectively. Moreover, $\gamma$ is the particle's Lorentz factor, so that $\gamma d\tau = dt$, where $\tau$ is the proper time along the path of the particle. Let us note that in the lowest nonrelativistic approximation $\gamma \approx 1$ and hence, $\boldsymbol{\omega}_{Thomas} \approx (\mathbf{a} \times \mathbf{v})/(2c^2)$. Thomas precession is ultimately due to the noncommutativity of Lorentz transformations. The spin is in general Fermi transported along the worldline of the spinning particle, namely, 
\begin{equation}\label{C2}
{\frac{dS^\mu}{d\tau}} + \Gamma^{\mu}_{\alpha \beta}u^{\alpha} S^{\beta} = {\frac {u^{\mu} a_{\nu}} {c^2}}S^{\nu}\,,
\end{equation}
where $a^\mu =Du^\mu/d\tau$ is the particle's 4-acceleration vector.

According to the quantum description of intrinsic spin, the particle Hamiltonian $H_0$ has,  in this case, an additional part given by $\boldsymbol{\omega}_{Thomas} \cdot \mathbf{S}$; that is, the Thomas precession of spin in the inertial frame implies that the Hamiltonian for the motion of the spinning particle is given by 
\begin{equation}\label{C3}
H = H_0 + \boldsymbol{\omega}_{Thomas}\cdot\mathbf{S}\,.
\end{equation}
Thomas precession and the spin-rotation coupling provide the necessary physical effects for a simple semiclassical description of spin motion in a rotating frame of reference~\cite{M2}. Recent work in this direction has focused on the physics of compound spin systems involving heavy ions in storage rings~\cite{Pav1,Pav2,L1, L2, L3}. 

It is interesting to note that Eq.~\eqref{C1} can be expressed as 
\begin{equation}\label{C4}
\boldsymbol{\omega}_{Thomas}=(\gamma - 1) {\frac{\mathbf{a} \times \mathbf{v}}{v^2}}\,,
\end{equation}
since $(\gamma + 1)(\gamma - 1)=\gamma^2 v^2/c^2$.  The acceleration vector of an arbitrary path in \emph{space} can be written as 
\begin{equation}\label{C5}
\mathbf{a} =  {\frac{dv}{dt}} \hat{\mathbf{t}} + {\frac{v^2}{\rho}} \hat{\mathbf{n}}\,, 
\end{equation}
where $\mathbf{v} = v \hat{\mathbf{t}}$ and $d\hat{\mathbf{t}}/dt = (v/\rho)\hat{\mathbf{n}}$. Here, the unit vectors $\hat{\mathbf{t}}$ and $\hat{\mathbf{n}}$ are respectively tangent and normal to the path and form the osculating plane, while $\rho$ is the instantaneous radius of curvature of the path. Thus the acceleration vector lies in the osculating plane and consists of the tangential acceleration $dv/dt$ and the centripetal acceleration $v^2/\rho$ that is directed toward the instantaneous center of curvature. The unit vectors $\hat{\mathbf{t}}$ and $\hat{\mathbf{n}}$ together with the binormal unit vector  $\hat{\mathbf{b}}=\hat{\mathbf{t}} \times \hat{\mathbf{n}}$ form the moving Frenet frame field along the path. The instantaneous angular velocity of the frame is given by  $\boldsymbol{\omega}(t) = (v/\rho)\hat{\mathbf{b}}$. It then follows from Eq.~\eqref{C5} that 
\begin{equation}\label{C6}
 \boldsymbol{\omega}_{Thomas} = -\,(\gamma -1)\boldsymbol{\omega}\,.
\end{equation}
In view of this result, namely, $\boldsymbol{\omega}_{Thomas}= \boldsymbol{\omega}-\gamma \boldsymbol{\omega}$, Thomas precession can be interpreted in terms of an overcompensation due to time dilation~\cite{MGL}. 

Let us now imagine that the acceleration of the spinning particle is due to gravity. In general relativity, the Newtonian concept of acceleration of gravity is nonexistent, as gravity is absorbed into the geometry of spacetime and  only nongravitational forces can be the source of true acceleration for a point particle. For the spin motion in Eq.~\eqref{C2}, the acceleration term on the right-hand side thus disappears and is replaced by the Christoffel term on the left-hand side. This is ultimately a consequence of Einstein's principle of equivalence and means that in Eq.~\eqref{C4}, we replace $\mathbf{a}$ by $-\mathbf{g}$ and $\gamma=dt/d\tau$ by the corresponding expression for proper time in the Schwarzschild geometry. The question is then the connection between this gravitational ``Thomas precession" and geodetic precession. This issue has been discussed in Ref.~\cite{MGL} and an exact correspondence has been pointed out for \emph{circular} orbits in the exterior Schwarzschild spacetime when the \emph{standard Schwarzschild coordinates} are employed---see pages 94 and 95 of Ref.~\cite{MGL}. On the other hand, we are interested here in the precession as perceived by our background observers and thus should compare the corresponding average precession rate with Eq.~\eqref{III14}. To order $c^{-2}$, we have $d\tau/dt=1-\phi-v^2/(2c^2)$; therefore, we find
\begin{equation}\label{C7}
 \left \langle -\,(\phi + {\frac{v^2}{2c^2}}){\frac{(\mathbf{g} \times \mathbf{v})}{v^2}}\right \rangle = \langle ^{(g)}\boldsymbol{\Omega}\rangle (1 - {\frac{1}{3}}e^2)\,,
\end{equation}
since
\begin{equation}\label{C8}
{\frac{1}{2\pi}} \int\limits_0^{2\pi} {\frac{(1 + e\cos\varphi)^2}{1+2e\cos\varphi + e^2}}~d\varphi = 1 - {\frac{1}{2}}e^2\,,
\end{equation}
which is valid for $e\in [0, 1]$. Thus to order $c^{-2}$, the average gravitational ``Thomas precession" in powers of the eccentricity $e$, $0 \le e<1$, is the same as the average  geodetic precession up to terms that are linear in the orbital eccentricity.

\end{document}